%% file: DKag.tex
\documentclass[aps,prl,twocolumn,superscriptaddress,merge,color]{revtex4-1}
\usepackage[pdftex]{graphicx}
\usepackage{float}
\usepackage{amsfonts}
\usepackage{textcomp}
\usepackage{color}
\usepackage{amsmath}
\usepackage[pdftex,colorlinks=true]{hyperref}
\usepackage{amssymb}
\usepackage{amsthm}
\usepackage{amsfonts}
\usepackage{bbm}
\usepackage{array}

\newcommand{\vect}[1]{\mathbf{#1}}

\newcommand{\bs}{{\bf{S}}}

\newcommand{\bL}{{\bf{L}}}

\newcommand{\ns}{N_{\mathrm{s}}}

\newcommand{\bea}{\begin{eqnarray}}
\newcommand{\eea}{\end{eqnarray}}

\newcommand{\beq}{\begin{equation}}
\newcommand{\eeq}{\end{equation}}

\def\sectionn#1{\noindent\underline{\it #1:}}

\def\bei{\begin{itemize}}
\def\eei{\end{itemize}}
\def\beq{\begin{equation}}
\def\eeq{\end{equation}}

\begin{document}

\title{Jammed spin liquid in the bond-disordered kagome Heisenberg
  antiferromagnet}

\author{Thomas Bilitewski} \affiliation{Max-Planck-Institut f\"{u}r Physik
  komplexer Systeme, N\"othnitzer Str.\ 38, 01187 Dresden, Germany}

\author{Mike E. Zhitomirsky} \affiliation{Universite Grenoble Alpes, CEA,
INAC-Pheliqs, 38000 Grenoble, France}

\author{Roderich Moessner} \affiliation{Max-Planck-Institut f\"{u}r Physik
  komplexer Systeme, N\"othnitzer Str.\ 38, 01187 Dresden, Germany}

\begin{abstract}
  We study a class of continuous spin models with bond disorder including the
  kagome Heisenberg antiferromagnet. For weak disorder strength, we find {\it discrete} ground states whose
  number grows exponentially with system size. These states do not exhibit
  zero-energy excitations characteristic of highly frustrated magnets but
  instead are local minima of the energy landscape.
  %, albeit with an anomalously soft excitation spectrum.
  This represents a spin liquid version of the
  phenomenon of jamming familiar from granular media and structural glasses.
  Correlations of this jammed spin liquid, which upon increasing the disorder
  strength gives way to a conventional spin glass, may be algebraic
  (Coulomb-type) or exponential.
\end{abstract}

\maketitle

\sectionn{Introduction} A large ground state degeneracy is a defining feature of
strong geometric frustration in classical spin systems. It underpins much of
their exotic properties, in particular the emergence of topological spin liquids
\cite{Moessner_2006}. For discrete spins
\cite{Wannier1950,AndersonFerrites1956}, the number of ground states can scale
exponentially in the system size, whereas for continuous spins the ground states
form a manifold whose dimension is proportional to the system size.
Comparatively little is known about the effect of disorder and lattice
distortions, with some pioneering works having unearthed both a capacity of spin
liquids to accommodate disorder
\cite{Shender_1993,MossnerBerlinsky1998,Henley2001}, and an immediate
instability towards spin glassiness for arbitrarily weak disorder
\cite{BellierHoldsworth_2001,Saunders_2007,Andreanov_2010,BergmanBalents,Wang_2007,Chern_2008,Roychowdhury2017}.
Given their large degeneracy, % of the ground state,
the geometrically
frustrated magnets should be particularly susceptible to perturbations and
disorder in the ideal structure. Such perturbations are necessarily present
in real materials and may themselves induce new
phenomena \cite{canalsoverview,Sala2014,Sen2015,Balents_2017}.
% Random strains in the material can via magneto-elastic coupling induce
% variations in the couplings which in the case of pyrochlore leads to a spin
% glass phase \cite{Saunders_2007}.
In addition, it has recently been realized that the field of classical spin
liquids may be richer than appreciated so far. New arrivals include
an anisotropic pyrochlore magnet exhibiting pinch-lines in the excitation
spectrum\cite{benton_pinchlines}, as well as spin liquids exhibiting
exponential, rather than Coulomb, correlations in the limit of low temperatures
\cite{rehn_newspinliquid_2017}.

Here, we present a family of continuous spin models which exhibit
a novel {\it jammed spin liquid} regime with an exponentially
large set of discrete ground states. We study in depth the kagome
Heisenberg magnet, where the jammed spin liquid appears most naturally. Like in
the clean system, the ground-state spin configurations minimize energy for every triangle,
but in contrast to it, they remain
disconnected exhibiting no non-trivial zero-energy modes. Still they show a softer
spectrum than that of a spin glass, which in turn appears at higher disorder strength.
%The ground states have a larger entropy, which we estimate as ${\cal S}\approx(\ln2)/3$,
%than the well-known coplanar kagome ground states --
%themselves unstable to bond disorder -- of the clean system.
Depending on model details,
the jammed spin liquid either inherits the algebraic
Coulomb correlations, or exhibits a disorder-screened version thereof.

Transitions in constraint satisfiability in {\it continuous} systems,
 at which an exponential number of {\it discrete} ground states
 appear, are known in the context of structural glasses and granular media under
 the heading of jamming \cite{Liu1998,LiuNagel2003,Liu2010}, from which we have
 borrowed the term. 
 Our model is a natural extension of these ideas to
 frustrated spin systems,  and we discuss possible interactions between these fields in the outlook.
 
Our analysis utilizes a number of different methods, including direct numerical
searches for ground states, and combining these with analytical continuity
arguments. In addition, we perform calculations within the self-consistent
Gaussian approximation (SCGA) \cite{Garanin_1996} to study the correlations on
large systems, as well as Monte-Carlo (MC) simulations to access finite
temperature properties and the spin glass phase. We finish with an outlook and a
discussion of connections to physics beyond spin systems.
 
\begin{figure}
  \begin{minipage}[t]{0.6\columnwidth}
  \vspace{0pt}   \includegraphics[width=.99\columnwidth]{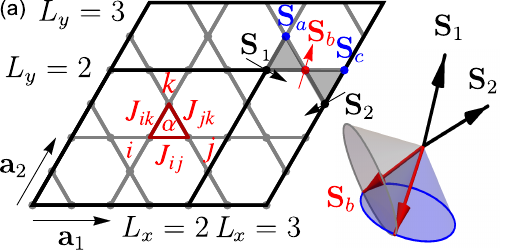}
  \end{minipage}
  \begin{minipage}[t]{0.38\columnwidth}
    \vspace{0pt} \includegraphics[width=.99\columnwidth]{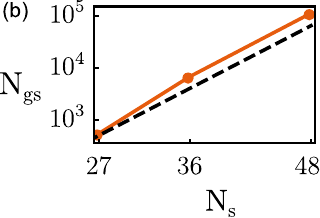}
  \end{minipage}
\caption{(a) Illustration of finite-size kagome lattices with primitive
  lattice vectors $\vect{a}_1=(1,0)$ and $\vect{a}_2=(1,\sqrt{3})/2$. Spins in
  triangle $\alpha$ at sites $i,j,k$ are coupled via $J_{ij}$.
  Black parallelograms demark the edges of square $L=L_x=L_y=2,3$ systems, for periodic boundary conditions opposite edges are identified.
  Marked spins/triangles pertaining to the transfer matrix construction (see text).
  (b) Number of ground states $N_{gs}$ vs.\ number of spins $\ns$
  compared with the scaling $N_{\mathrm{gs}} \sim 2^{\ns/3}$ (dashed line).
  \label{fig:illustration_lattices_and_construction}
}
\end{figure}

\sectionn{Model Hamiltonian} Our starting point is the classical Heisenberg
model of $O(3)$ spins on a kagome lattice
\bea
H&=&\sum_{\langle ij\rangle}
J_{ij} \, \bs_i\cdot\bs_j \ ,%\\
\label{eq:KAFM}
\eea
with random nearest-neighbor exchanges $J_{ij}>0$.
For every triangle $\alpha$ formed by sites $ijk$,
see Fig.~\ref{fig:illustration_lattices_and_construction}(a),
we define
$\gamma_{i\alpha}=\sqrt{{J_{ij}J_{ik}}/{J_{jk}}}$ and rewrite $H$ as
\begin{equation}
  H=\frac{1}{2} \sum_\alpha \bL_\alpha^2  ,\quad \text{with} \quad \bL_\alpha=\sum_{i\in\alpha}\gamma_{i\alpha}\bs_i \, ,
  \label{eq:H_general}
\end{equation}
up to a constant energy shift. The above form provides a set of local ground-state constraints,
which may or may not be satisfied simultaneously, see below.
Inversely, Eq.~(\ref{eq:H_general}) generates a
bond-disordered model (\ref{eq:KAFM}) with couplings $J_{ij} =
\gamma_{i\alpha} \gamma_{j\alpha}$ between spins $ij$ in triangle $\alpha$.
% \begin{figure}
%   \begin{minipage}{0.49\columnwidth}
%     \includegraphics[width=.99\columnwidth]{illustration_lattices.pdf}
%   \end{minipage}
%    \caption{Illustration of finite-size kagome lattices with primitive lattice
%      vectors $\vect{a}_1=(1,0)$ and $\vect{a}_2=(1,\sqrt{3})/2$. We also
%      indicate one triangle $\alpha$ composed of sites $i,j,k$ with bond-couplings $J_{ij}$.
%      Black parallelograms demark the edges of $L_x=L_y=2,3$ systems and for periodic boundary conditions opposite edges are identified.
%      \label{fig:lattices}}
%    \end{figure}

More generally, the Hamiltonian (\ref{eq:H_general}) can be defined for
$O(n)$ spins on frustrated lattices consisting of fully-connected units/simplices of
$q$ spins: $n=2,3$ for $XY$ and Heisenberg spins, whereas
$q=3,4$ for triangles and tetrahedra. For $\gamma_{i\alpha}\equiv 1$, one recovers
the regular frustrated spin models for a range of lattices, such as kagome, checkerboard,
pyrochlore or maximally frustrated honeycomb models, all of which exhibit order by disorder (obdo) for small
$n$ \cite{Chalker_1992,ChandraColemanRitchey1993,HarrisBerlinsky1992,Moessner_1998,*MoessnerPyro1998,RehnHoneycomb2016}
and a classical spin liquid for large $n$
\cite{Garanin_1999,Moessner_1998,*MoessnerPyro1998}. For $q>3$,
a bond-disordered model (\ref{eq:KAFM}) generated by (\ref{eq:H_general}) has correlations
between bond amplitudes in the same simplex. For the kagome lattice ($q=3$) the
mapping between $J_{ij}$ and $\gamma_{i \alpha}$ is one-to-one, and therefore we
focus on this example as the most natural one (for the $XY$ spins on the checkerboard
lattice see \cite{supplemental}).

We identify and study two classes of distributions of $J_{ij}$ on the kagome
lattice with somewhat different phenomenology. The first one is the bond-disorder model (BDM)
with $J_{ij}$ chosen uniformly within
$(1-\delta,1+\delta)$ and $\delta<1$. The second class called the
`maximally disordered Coulomb model' (MCM) is constructed from
Eq.~(\ref{eq:H_general}) in the following way: we set $\gamma_{i\alpha}=\gamma_i
\gamma_{\alpha}$ assigning a factor $\gamma_{i}$ to every spin and an
additional factor $\gamma_{\alpha}$ to every triangle.
Both are chosen uniformly within $(1-\delta,1+\delta)$ which ensures
that the models have the same critical point $\delta_c$
\cite{supplemental}). The MCM is defined by $5 L^2$ random parameters as opposed
to $6 L^2$ for the BDM. We believe
that MCM is the most general model preserving the Coulomb correlations; in
particular, it saturates the number of degrees of freedom allowed accounting for
$L^2$ ``star-conditions'' very recently identified in
Ref.~\cite{Roychowdhury2017}.

\sectionn{Ground state construction and counting}
The lowest-energy classical spin configurations must
satisfy the set of local constraints
\begin{equation}
  \bL_{\alpha}=0 \ \ \ \forall \alpha \ .
  \label{eq:constraintp}
\end{equation}
For an isolated triangle $\alpha$, the constraint implies that three vectors $\gamma_{i\alpha}{\bf S}_i$
form a closed triangle in spin space with side lengths $\gamma_{i\alpha}$.
This can be always achieved once the corresponding $\gamma_{i\alpha}$ obey the triangle inequality,
which in turn restricts $\delta\leq 1/3$ \cite{supplemental}.
For the full lattice, an indicator of the dimensionality of the ground state manifold is given by
the Maxwellian counting argument which compares the number of the degrees of freedom $D$
to the number of ground state constraints $K$.
Similar to the regular kagome Heisenberg antiferromagnet \cite{Moessner_1998},
$D=K$ in our case suggesting no degeneracy.
However, the constraint counting does not account for possibly dependent or inconsistent
constraints. Next, we give an explicit
construction, which shows that ground states of the full system are generically
discrete and provides an estimate of their number.

% \begin{figure}
%   \begin{minipage}{0.44\columnwidth}
%     \includegraphics[width=.99\columnwidth]{{illustration_ground_state_construction}.pdf}
%   \end{minipage}
%   \begin{minipage}{0.44\columnwidth}
%     \includegraphics[width=.99\columnwidth]{{number_of_states}.pdf}
%   \end{minipage}
%   \caption{Left: Kagome lattice and illustration of the ground state
%     construction.
%     %The gray and black spins are known. We next determine the three spins within the
%     %gray ellipses.
%    Right: Number of ground states $N_{gs}$ vs.\ number of spins $\ns$ compared with
%    the scaling $N_{\mathrm{gs}} \sim 2^{\ns/3}$ (dashed line).
%    \label{fig:illustration_groundstate_construction}}
%  \end{figure}
To construct the ground states in the bulk, {\it i.e.}, ignoring the boundaries,
we proceed from layer to layer in the spirit of a transfer matrix. 
Let us consider the group of three spins $\mathbf{S}_a$, $\mathbf{S}_b$, $\mathbf{S}_c$, see
Fig.~\ref{fig:illustration_lattices_and_construction}, such that
all spins in the lower layer and to the left of the group are already fixed.
The three spins belong to a pair of up/down triangles of the lattice.
In each of the two triangles one spin
$\mathbf{S}_{1(2)}$ is fixed and one unknown spin
$\mathbf{S}_b$ (red) is shared between the up and down triangles. The ground
state constraint determines angles between spins in a triangle, the only remaining
freedom is rotation of the undetermined spins around the fixed spin
$\mathbf{S}_{1(2)}$. This rotation makes the common spin $\mathbf{S}_b$ sweep
out two distinct conic sections (whose opening angle depends on the random bond
couplings) of the unit sphere as shown in
Fig.~\ref{fig:illustration_lattices_and_construction}, which generically have
either none or two points of intersection. When there are two intersection
points, there is a {\it discrete} choice between them, yielding an orientation
of $\mathbf{S}_b$ consistent with the constraints in both triangles. This step
is then repeated to determine all spins throughout the lattice.

Ignoring the possibility of inconsistent constraints that yield no crossing,
the above procedure estimates the
number of ground states as $N_{\mathrm{gs}}\sim 2^{N_s/3}$ for
$N_s$ spins in the lattice in accordance with the number of up-down triangle pairs.
Interestingly, the corresponding entropy ${\cal S}/N_s\approx \frac{1}{3}\ln 2$ is larger 
than the entropy of the well-known coplanar
states of the clean system ${\cal S}/N_s \approx\ln(1.13)$ \cite{Baxter_1970}.
Figure~\ref{fig:illustration_lattices_and_construction}(b) shows enumeration
results on small finite systems consistent with the derived scaling.
We also provide arguments for the continuity of each state as a function of
$\delta$ in the Supplemental Material \cite{supplemental}.

\sectionn{Correlations}
\begin{figure}
  \begin{minipage}[t]{.49\columnwidth}
   \vspace{0pt}
    \includegraphics[width=.99\columnwidth]{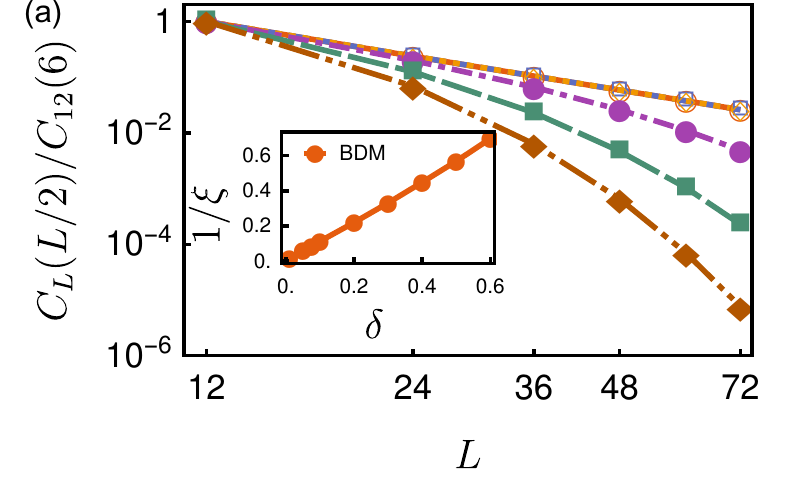}
  \end{minipage}
  \begin{minipage}[t]{0.49\columnwidth}
   \vspace{0pt}
    \includegraphics[width=.99\columnwidth]{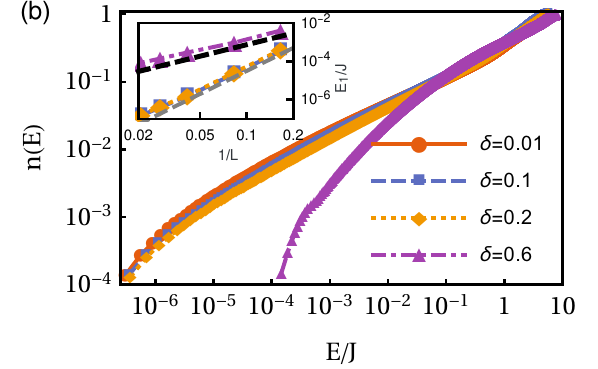}
  \end{minipage}
\caption{
   (a) Finite-size scaling of $C_L(L/2)=\langle \vect{S}(0) \cdot \vect{S}(L/2)
   \rangle$ %with linear system size $L$ %normalised to their value at $L=12$
   on a log-scale for MCM (open symbols) and
   BDM (filled symbols) for $\delta=0.1,0.2,0.3$ (circles, squares, diamonds) on
   $L\times L$ systems.
   Inset: Inverse of the correlation length $1/\xi$ as a function of $\delta$
   obtained from the long distance behaviour, $\langle \vect{S}(0) \cdot \vect{S}(r)
   \rangle \sim e^{-r/\xi}$, of the BDM on a $144\times36$ system.
(b) Cumulative density of the eigenvalues of the Hessian matrix for the jammed spin liquid $\delta<1/3$ and spin glass.
  Inset: Scaling of the lowest (non-trivial) eigenvalue $E_1$
  shows the relative softness of the jammed spin liquid. Dashed
  lines are guides to the eye with $L^{-2}$ (black) and $L^{-4}$ (gray).
  \label{fig:cors_and_soft_modes}}
\end{figure}
% \begin{figure}
%   \begin{minipage}{\columnwidth}
%     \includegraphics[width=.99\columnwidth]{fig_cors_finite_size_scaling_all.pdf}
%   \end{minipage}
%  \caption{
%    Finite-size scaling of $C_L(L/2)=\langle \vect{S}(0) \cdot \vect{S}(L/2)
%    \rangle$ %with linear system size $L$ %normalised to their value at $L=12$
%    on a log-scale for MCM (open symbols) and
%    BDM (filled symbols) for $\delta=0.1,0.2,0.3$ (circles, squares, diamonds) on
%     $L\times L$ systems.
%    Inset: Inverse of the correlation length $1/\xi$ as a function of $\delta$
%    obtained from the long distance behaviour, $\langle \vect{S}(0) \cdot \vect{S}(r)
%    \rangle \sim e^{-r/\xi}$, of the BDM on a $144\times36$ system.
%    \label{fig:cors}}
%  \end{figure}
We compute the correlations within the SCGA which is exact in the limit of spin components $n \rightarrow \infty$
\cite{Stanley_1968}, and provides quantitative results for the low temperature
correlations at finite $n$ \cite{Garanin_1999}. It allows to access considerably
larger systems than with explicit energy minimisation; where both are possible,
the results agree with each other and with our Monte Carlo simulations
\cite{supplemental}.

There is a fundamental difference between BDM and MCM, as displayed in
Fig.~\ref{fig:cors_and_soft_modes}(a), which shows the finite-size scaling
analysis of $C_L(L/2)= \langle \vect{S}(0) \cdot \vect{S}(L/2) \rangle$. The MCM
retains the algebraic correlations characteristic of the Coulomb phase present
at large-$n$, but does not exhibit the peaks present for the disorder-free case
of $n=3$ resulting from order by disorder. By contrast, the BDM finds a
crossover to exponential decay with a correlation length $\xi\sim1/\delta$
(inset of Fig.~\ref{fig:cors_and_soft_modes}(a)). This follows from the fact
that the MCM straightforwardly permits the definition of a height-model (which
implies the $L^{-2}$ behaviour) analogous to the disorder-free case
\cite{Huse_1992}, whereas in the BDM this appears to be impossible. The screened
correlations of the BDM are comparable to those of the clean system at a
temperature $T^{*}\sim \delta^2 $ (suppl. mat. \cite{supplemental}).

\sectionn{Low energy spectrum of Hessian} We study the quadratic energy cost of
deformations of the ground states via the spectrum of the Hessian matrix.
Importantly, we find no zero-modes for either the BDM or the MCM. This is in
stark contrast to the coplanar states of the clean kagome system which have an
extensive number of exact zero-modes. In the language of mechanical lattices our
spin ground states are fully rigid \cite{Kane_2013,Lawler_2016}.

% \begin{figure}
%   \begin{minipage}{0.99\columnwidth}
%     \includegraphics[width=.99\columnwidth]{soft_modes.pdf}
%   \end{minipage}
% \caption{Cumulative density of the eigenvalues of the Hessian matrix for the jammed spin liquid $\delta<1/3$ and spin glass.
%   Inset: Scaling of the lowest (non-trivial) eigenvalue $E_1$
%   shows the relative softness of the jammed spin liquid. Dashed
%   lines are guides to the eye with $L^{-2}$ (black) and $L^{-4}$ (gray).
%   \label{fig:soft_modes}}
% \end{figure}

Nonetheless, we find that the excitations in the jammed spin-liquid are
considerably softer than in a spin glass, see
Fig.~\ref{fig:cors_and_soft_modes}(b), e.g. in that the smallest eigenvalue of
the Hessian spectrum vanishes with a higher power of system size. We also note
that at small $\delta$, the spectrum appears to become independent of $\delta$,
suggesting it also describes the excitations of the discrete noncoplanar ground
states of the disorder-free system.

\begin{figure}
  \begin{minipage}{0.99\columnwidth}
    \includegraphics[width=.99\columnwidth]{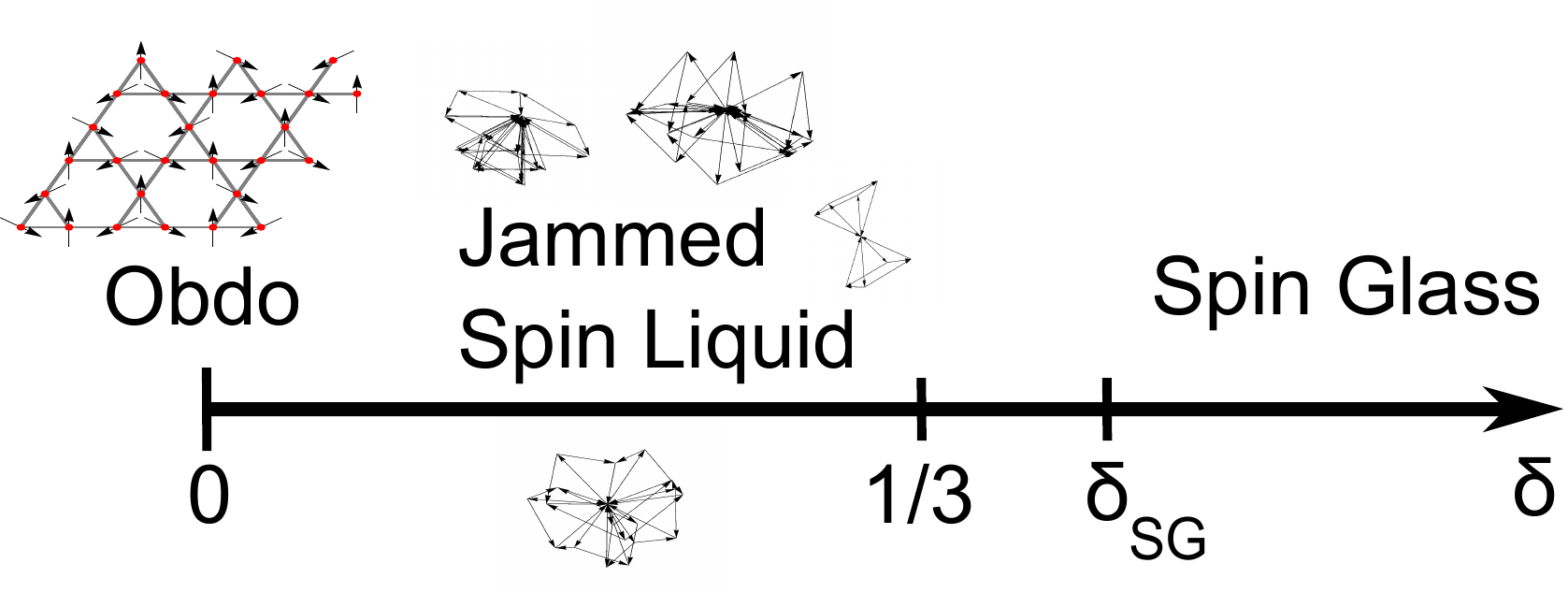}
  \end{minipage}
\caption{Phase Diagram as a function of disorder strength, with the jammed
spin liquid appearing for $0<\delta<1/3$. Illustrations are of the
$\sqrt{3}\times \sqrt{3}$ state and of non-coplanar finite size jammed
spin-liquid states. The precise value of $\delta_{\mathrm{SG}} \geq \delta_c$ is not known.
\label{fig:phase_diagram}}
\end{figure}

\sectionn{Phase Diagram (Fig.~\ref{fig:phase_diagram})}
The jammed spin liquid is terminated by two different states for low and high
$\delta$. We consider these in turn.

\begin{figure}
  \begin{minipage}{\columnwidth}
    \includegraphics[width=.99\columnwidth]{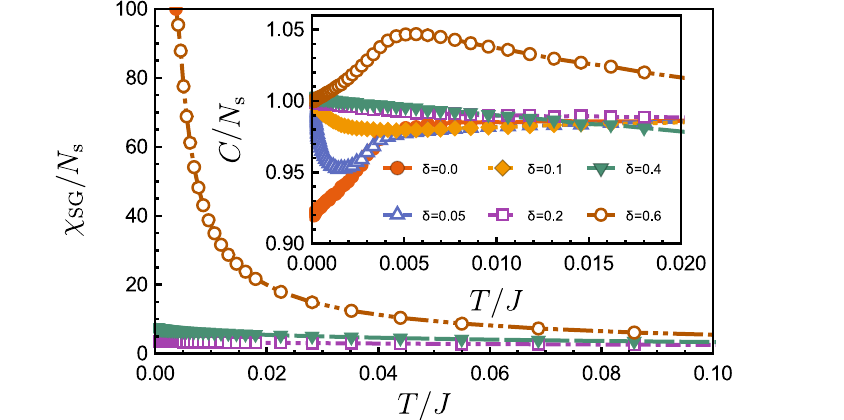}
  \end{minipage}
    \caption{Spin glass susceptibility $\chi_{\mathrm{SG}}$ vs. temperature in units
      of the bond strength $T/J$ from MC simulations for the BDM.
     Inset: Specific heat per spin $C/\ns$ vs. temperature showing absence of
     order by disoder for $\delta>0$.
     \label{fig:MC_results}}
 \end{figure}

The clean system, $\delta=0$, is the archetypal frustrated magnet exhibiting
order by disorder in the form of coplanar states
\cite{Chalker_1992,ChandraColemanRitchey1993,HarrisBerlinsky1992,Huse_1992,Zhitomirsky_2008}
with weak $\sqrt{3} \times \sqrt{3}$ translational symmetry breaking
\cite{Zhitomirsky_2008,ChernMoessner2012}. Bond disorder is inconsistent with
coplanarity in the sense that the energy of coplanar states exceeds that of the
non-coplanar ground states (suppl. mat. \cite{supplemental}). Since the order by
disorder is driven by excess soft modes, it can be diagnosed by their signature
in the reduced heat capacity \cite{Chalker_1992}. The heat capacity, computed
from fluctuations of the internal energy in the MC simulations as $C= \left(
  \langle E^2 \rangle -\langle E \rangle^2\right)/T^2$, is shown in the inset of
Fig.~\ref{fig:MC_results}. The disorder-free value of $C=11/12$ at low
temperatures \cite{Chalker_1992,Zhitomirsky_2008} is replaced by $C=1$ for any
$\delta>0$ consistent with our finding that there is no extensive number of soft
modes.
In addition, we clearly observe three distinct signatures in the heat capacity,
a ``dip'' in the JSL phase ($\delta=0.1$), an intermediate regime with flat
behaviour ($\delta=0.4$), and a ''bump`` in the spin glass phase ($\delta=0.6$).

\begin{figure}
  \begin{minipage}{0.99\columnwidth}
    \includegraphics[width=.99\columnwidth]{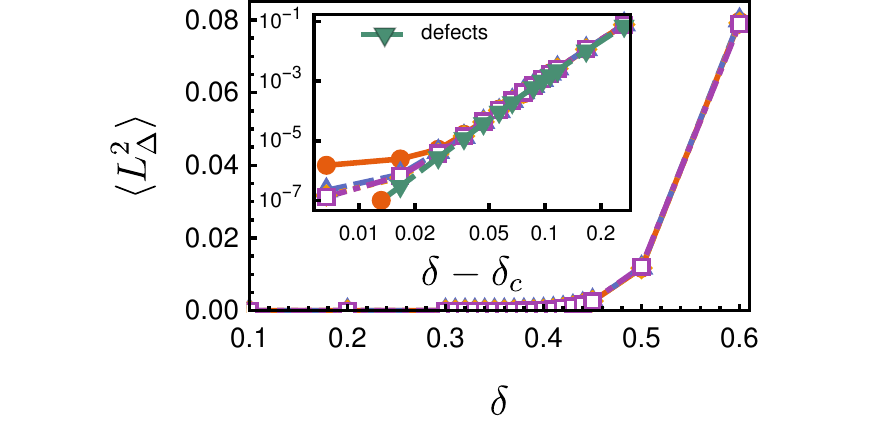}
  \end{minipage}
 \caption{Residual energy per triangle of the BDM vs.\ $\delta$ for $L=12,24, 36, 48$ (circles, triangles, diamonds, squares).
  % Statistical errors are smaller than symbol size.
   Inset: Same on a log-scale vs. $\delta-\delta_c$ consistent with
   $\left<\vect{L}_{\Delta}^2\right> \sim (\delta-\delta_c)^5$%
   % \sim \frac{27}{5} \left[ \frac{3}{2} \left(\delta-\delta_c \right)\right]^5$
   and from a direct evaluation of the energy of ``defect'' triangles with
   collinear spins.
   \label{fig:gs_L0}}
\end{figure}

Throughout the jammed spin liquid, the ground state constraints are obeyed,
exactly for the MCM (suppl. mat \cite{supplemental}); for the BDM, there is one
unsatisfiable global constraint imposed by the periodic boundary conditions. The
total sum of spins on the up and down triangles has to be equal, as this just
amounts to a different bookkeeping of all the spins in the system. Indeed, we
find that (Fig.~\ref{fig:gs_L0}) for $\delta<\delta_c$, $\left<
  \vect{L}_{\Delta}^2\right> \sim L^{-4}$ vanishes in the thermodynamic
limit, %only depends very weakly on $\delta$,
consistent with a single (or more generally a non-extensive number of)
unsatisfiable constraints distributed over $N_{\Delta} \sim L^2$
triangles. %yielding a scaling of the residual energy per triangle $\sim N_{\Delta}^{-2} \sim L^{-4}$.
% For the MCM (not shown) the residual energy is strictly zero below the
% transition.

The jammed spin liquid regime terminates at a critical disorder strength
$\delta_c=1/3$. This threshold is related to the impossibility to satisfy the
local constraint $\bL_{\alpha}=0$ for a large disparity between bond values.
Beyond $\delta_c$ coupling configurations appear for which the
groundstate constraint cannot be satisfied (suppl.mat. \cite{supplemental}).
% This threshold is related to the impossibility to satisfy the
% local constraint $\bL_{\alpha}=0$ for a large disparity between bond values. At
% $\delta_c=1/3$, individual triangles start exhibiting collinear spin
% configurations with a pair of parallel spins antialigned with the third, if the
% weakest bond is $J_{ij}=1-\delta_c$ and the other two
% $J_{ik}=J_{jk}=1+\delta_c$. Beyond this, $\bL_{\alpha}=0$ becomes unsatisfiable.
Near $\delta_c=1/3$, the probability of choosing such bond couplings
$\{J_{ij}\}$ grows as $(\delta-\delta_c)^3$. Together with
$\bL_{\Delta} \sim (\delta-\delta_c)$ this yields
$\left<\vect{L}_{\Delta}^2\right> \sim \delta^{5}$, in agreement with an
analysis of a single triangle, $\left<\vect{L}_{\Delta}^2\right> \sim
\frac{27}{5}\left[ \frac{3}{2}\left(\delta-\delta_c \right)\right]^5 $.
% For the MCM the residual energy scales analogously above the transition.

On further increasing $\delta$, the system turns into a conventional spin glass
beyond a $\delta_{\mathrm{SG}} \geq \delta_c$ as evidenced by the diverging
spin-glass susceptibility (Fig.~\ref{fig:MC_results}). In the jammed spin-liquid
$\delta<\delta_c$ and for intermediate $\delta_C<\delta<\delta_{\mathrm{SG}}$,
$\chi_{\mathrm{SG}}(T)$ remains flat down to the lowest temperatures.

\sectionn{Open questions and connections} A number of questions follow naturally
from this study, e.g. whether the jammed spin liquid entropy may be determined
exactly. Also, its low-temperature dynamical properties should be worth
investigating, as it appears to fit neither previous examples of conventional
spin liquids or kagome Heisenberg magnets
\cite{Moessner_1998,*MoessnerPyro1998,Conlon2009,Robert2008,Taillefumier2014,Keren1995},
nor a Halperin-Saslow picture of a disordered magnet with a finite spin
stiffness \cite{HalperinSaslow1977}.

 Regarding the phase diagram, we have not numerically determined the exact
 point, $\delta_{SG}\geq\delta_c$ of the spin glass transition. The possibility
 of another new regime for $\delta_c<\delta<\delta_{SG}$ appears not unnatural
 given, (i) that the excess energy for $\delta>\delta_c$ can be captured by just
 counting the number of triangles with collinear spins, without taking into
 account any collective physics between them, (ii) neither $C$
 (Fig~\ref{fig:MC_results}) nor $\chi_{\mathrm{SG}}$ (suppl. mat.
 \cite{supplemental}) show within our numerical precision indications of
 non-analytical behaviour even at $\delta=0.4$, (iii) the capacity of spin
 liquids to screen disorder \cite{Rehn2015}.

 Potential candidate materials are the kagome hydronium jarosites
 \cite{Wills1996,Wills2008}, which even though showing a spin freezing
 transition \cite{Wills2000}, appear to be different from
 conventional spin glasses \cite{Wills2000a,Wills2002,Ladieu2004}. 
 Besides disorder and geometric frustration, the anisotropic distortion may cause
 glassy behavior in these systems \cite{Wills2008}, and it is an interesting question whether
 anisotropy or second-neighbor interactions would stabilize the glassy
 phase in the model studied here. In contrast to the pyrochlore lattice where
 infinitesimal disorder produces a spin-glass transition
 \cite{Saunders_2007}, we find the spin liquid to be
 stable for the kagome-lattice model.  This implies that for antiferromagnetic materials 
 with the kagome structure, 
 weak disorder alone cannot account for the observed spin freezing.

 The termination of the jammed spin liquid is of broader interest on account of
 its connection to other fields of statistical mechanics. The marginality of the
 kagome Heisenberg magnet in Maxwellian constraint counting was noted already a
 long time back \cite{Moessner_1998,*MoessnerPyro1998}, when it was also
 realized that such marginal constraint tends to underpin the order by disorder
 phenomenon, which we have here found is in turn suppressed by bond disorder.
 Recent developments have emphasized connections to a broader class of systems,
 in particular mechanical Maxwell lattices \cite{Kane_2013}, with implications
 for topological aspects of the excitation spectrum, and the local stability of
 distorted kagome ground states \cite{Lawler_2016,Roychowdhury2017}, which may
 be of relevance to our numerics on the Hessian matrix.

Investigating the connection of our spin model to `conventional' 
jamming \cite{Liu1998,LiuNagel2003,Liu2010}, 
 will be a most interesting topic
 for future study. Dynamical and nonlinear
 properties should be of particular interest, see e.g. 
 the very recent preprints \cite{Lubchenko2017,Wu2017}. 
 Crucially, there are some properties specific to our
 setting, including an extended and stable jammed regime; the gradual onset
 and spatial localization of the added constraints; the possibility of the
 satisfiable regions of the system acting as a medium generating effective
 interactions between the latter; and the peculiar onset of the nonzero energy
 density, which in turn will depend on details of the disorder distribution.

 This set of questions has also been given concerted attention in a more
 "computer-science'' context \cite{Franz2017}, where the low (high)-disorder
 regime corresponds to the (UN)SAT regime of a constraint-satisfaction problem.
 [In passing, we note that in the language of that community, the SAT/jammed
 spin liquid regime is unfrustrated, as all terms in the Hamiltonian can be
 simultaneously satisfied.] We hope our work will stimulate work establishing
 connections between all of these topics.

 \sectionn{Acknowledgements} We thank J. T. Chalker, Chris Laumann, A.
 Scardicchio for helpful discussions, and V. Vitelli for pointing us in the
 direction of the jamming phenomena. This work was in part supported by Deutsche
 Forschungsgemeinschaft via SFB 1143.

\bibliographystyle{plain}
%\bibliography{kagome_bib}{}
\input{DKag.bbl}

%%%%%%%%%% Merge with supplemental materials %%%%%%%%%%
\clearpage

\begin{center}
\textbf{\large Supplemental Material: Jammed spin liquid  in the bond-disordered kagome Heisenberg antiferromagnet}
\end{center}
%%%%%%%%%% Merge with supplemental materials %%%%%%%%%%
%%%%%%%%%% Prefix a "S" to all equations, figures, tables and reset the counter %%%%%%%%%%
\setcounter{equation}{0}
\setcounter{figure}{0}
\setcounter{table}{0}
\setcounter{page}{1}
\makeatletter
\renewcommand{\theequation}{S\arabic{equation}}
\renewcommand{\thefigure}{S\arabic{figure}}
\renewcommand{\bibnumfmt}[1]{[S#1]}
\renewcommand{\citenumfont}[1]{S#1}
%%%%%%%%%% Prefix a "S" to all equations, figures, tables and reset the counter %%%%%%%%%%

\section{Kagome model}
\sectionn{Analysis of a single triangle}
For both the BDM and the MCM models the critical disorder strength $\delta_c$ is related to
the impossibility to satisfy the local constraint $L_{\alpha} = \sum_{i \alpha}
\gamma_{i \alpha} \vect{S}_{i}=0$. Geometrically, this means that the spins form
a closed triangle with side lengths $\gamma_{i \alpha}$. Thus, there is no
possible solution if any side is larger than the sum of the other two $\gamma_1
> \gamma_2 + \gamma_3$. We also note that this shows that the minimal energy in an isolated
triangle is $0$ if $\gamma_1 \le \gamma_2 + \gamma_3$ and
$\gamma_1-(\gamma_2+\gamma_3)$ if $\gamma_1 > \gamma_2+\gamma_3$ (where we assume them to be ordered
in decreasing magnitude).
The critical point occurs exactly when side lengths that satisfy $\gamma_1 = \gamma_2 +
\gamma_3$ become possible, the triangle in spin space becomes a collinear configuration with two
parallel spins with small side lengths anti-parallel to a third spin with a
large side length.
 Below we will refer to triangles that do not allow a
zero-energy solution as ``defect'' triangles.

\begin{figure}
\begin{minipage}{0.99\columnwidth}
  \centering
  \includegraphics[width=.99\linewidth]{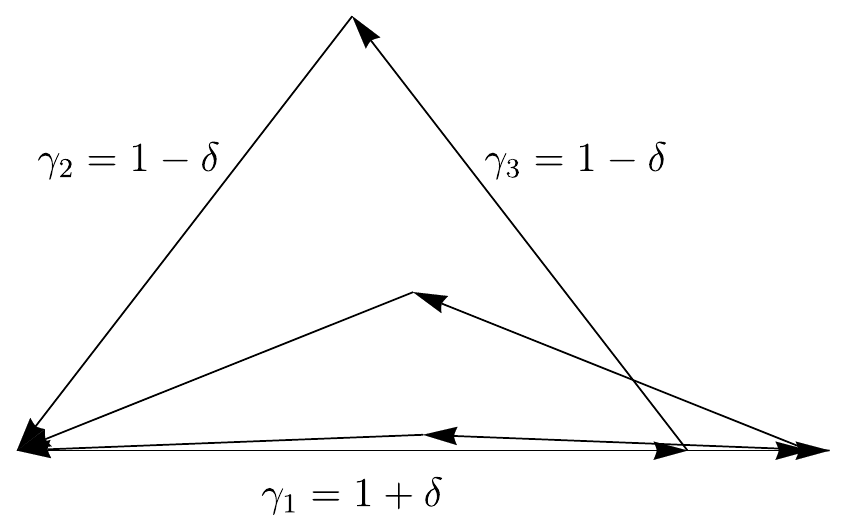}
\end{minipage}
\caption{The groundstate constraint $\gamma_1 \vect{S}_1+\gamma_2 \vect{S}_2
  +\gamma_3 \vect{S}_3=0$ implies that the spins form a closed triangle. Here
  illustrated for $\gamma_1 = 1+\delta$, $\gamma_2=\gamma_3=1-\delta$ with $\delta=0.1, 0.3, 0.333$.
  \label{fig:illustration_single_triangle}}
\end{figure}

We illustrate this transition in Fig.~\ref{fig:illustration_single_triangle} for
increasing disparity between two small and one large side length as relevant to the transitions in the
BDM and MCM.
We show below that for the choices of the couplings made in the main text, both
models have the same critical point $\delta_c =1/3$.

For the MCM we have $\gamma_{i \alpha} = \gamma_i \gamma_{\alpha}$. For the
groundstate constraint $\gamma_{\alpha}$ does not matter as it is a common factor
for all spins in a triangle.
As we choose $\gamma_i$ in $(1-\delta,1+\delta)$ we have at the critical point
\begin{align}
   (1+\delta_c) &= 2 \, (1-\delta_c)\, ,
\end{align}
thus $\delta_c^{\textrm{MCM}}=1/3$.

For the BDM we have $\gamma_{i \alpha} = \sqrt{\frac{J_{ij} J_{ik}}{J_{jk}}}$
and $J_{ij}$ in $(1-\delta,1+\delta)$. To obtain two minimal scaling factors and
one maximal scaling factor the couplings need to be $J_{ij}=1-\delta$ and
$J_{jk}=J_{ik}=1+\delta$.
Then we have at the critical point
\begin{align}
  \sqrt{\frac{(1+\delta_c^{\textrm{BDM}})(1+\delta_c^{\textrm{BDM}})}{1-\delta_c^{\textrm{BDM}}}} &= 2 \,
    \sqrt{\frac{(1+\delta_c)(1-\delta_c)}{1+\delta_c}} \, ,
\end{align}
thus, $\delta_c^{\textrm{BDM}}=1/3$.

We emphasise that this reasoning is based on the study of a single isolated
triangle. The fact that the model on the full connected kagome lattice shows the transition at
the same point is non-trivial.

% \sectionn{Conventions for finite lattices}
% We simulate the system on lattices in the shape of a parallelogram with edges
% parallel to the primitive basis vectors $a_1$ and $a_2$ of the kagome lattice.
% We specify systems by their linear dimensions $L_x$ and $L_y$ in the
% directions $a_1$ and $a_2$, and for ``square'' systems we only specify
% $L=L_x=L_y$. The number of spins is $N_s = 3 L_x L_y$.

% Periodic boundary conditions are enforced by identifying opposite edges.
% We show an illustration of these finite-size kagome lattices in
% Fig.~\ref{fig:lattices}.

% \begin{figure}
% \begin{minipage}{0.99\columnwidth}
%   \includegraphics[width=.99\columnwidth]{illustration_lattices.pdf}
% \end{minipage}
%    \caption{Illustration of finite-size kagome lattices with the primitive lattice vectors
%      $\vect{a}_1=(1,0)$ and $\vect{a}_2=1/2(1,\sqrt{3})$. The black
%      parallelograms demark the edges of $L_x=L_y=2,3,4$ systems.
%  \label{fig:lattices}}
% \end{figure}

\sectionn{Counting ground states}
We numerically search for ground states of the MCM
with a fixed disorder realisation at disorder
strength $\delta=0.1$.
We perform this search on periodic clusters of linear dimensions
$(L_x,L_y)=(2,2),(3,3),(4,3),(4,4)$ obtaining $10^3,10^4,10^5,10^6$ ground states.

For each of these states we compute the spectrum of its Hessian.
We then classify the ground states into distinct groups according to the first 10 eigenvalues of the
spectrum.
In table~\ref{tab:no_of_states} we summarise the results of these enumeration searches.
\begin{table}
\centering
\begin{tabular}{c r r r} \toprule
 $L_x,L_y$ & Samples & $2^{\ns/3}$  & $N_{\mathrm{gs}}$  \\
 2,2       & $10^3$ & 16                 & 4     \\
 3,3       & $10^4$ & 512                & 558   \\
 4,3       & $10^5$ & 4096               & 6910 \\
 4,4       & $10^6$ & 65536              & 113899\\
% \bottomrule
\end{tabular}
\caption{Results of the enumeration search.
  System sizes,
  Number of ground state searches,
  $2^{\ns/3} $expected number of ground states based on the bulk scaling,
  $N_{\mathrm{gs}}$ number of distinct ground states found.\label{tab:no_of_states}}
\end{table}
%
%\begin{table}
% \centering
% \begin{tabular}{c r r r r r} \toprule
%  $L_x,L_y$ & Samples & $N_{\mathrm{exp}}$  & $N_{\mathrm{exp}} \log(N_{\mathrm{exp}})$ & $N_{\mathrm{st}}$ & $N_{\mathrm{est}}$ \\
%  2,2       &$10^3$ & 16            & 44    & 4       &  4        \\
%  3,3       &$10^4$ & 512           & 3194  & 558     &  558        \\
%  4,3       &$10^5$ & 4096          & 34069 & 6910     &  \\
%  4,4       &$10^6$ & 65536         & 726817& 113899  &  121474 $\pm$ 115\\
% % \bottomrule
% \end{tabular}
% \caption{Results of the enumeration search.
%   System sizes,
%   Number of ground state searches,
%   $N_{\mathrm{exp}}$ expected number of ground states based on the bulk scaling
%   $N_{\mathrm{exp}} \sim 2^{N_{\mathrm{s}}/3}$,
%   expected number of samples required to find all these states for equally
%   likely states,
%   $N_{\mathrm{st}}$ number of distinct ground states found,
%   $N_{\mathrm{est}}$ estimated number of distinct ground states based on a
%   poisson fit to the state frequencies in the ground state search \label{tab:no_of_states}}
% \end{table}

\begin{figure}
\begin{minipage}{0.99\columnwidth}
  \centering
  \includegraphics[width=.99\linewidth]{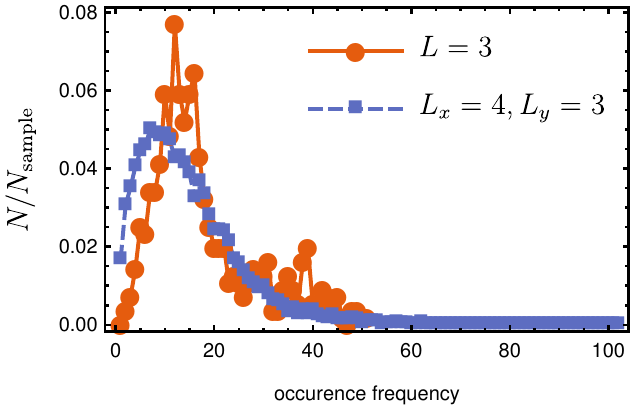}
\end{minipage}
\caption{Number of ground states $N$ that occur with given frequency in our
  ground state search normalised to total number of samples.
\label{fig:states_search_frequencies}}
\end{figure}

Characteristic distributions of the frequency counts, i.e. the probabilities
that a ground state occurs a certain number of times in our search are
shown in Fig.~\ref{fig:states_search_frequencies}, which permit an estimate of
the number of ground states missed by the search by fitting to a Poissonian distribution.

\sectionn{State continuity and fidelity}
Here, we consider the evolution of the classical ground states with
disorder strength and their connection to the states of the clean model via the
state fidelity.

To this end we fix a particular disorder realisation $J_{ij}=1 + \delta_{ij}$
with $\delta_{ij}$ uniform in $(-1,1)$ and then rescale the random part
$\delta_{ij}$ by the disorder strength $\delta$.
We then define the fidelity $F$ between states for a fixed disorder
realisation at different disorder strengths as $F(\delta,h)= \left|  \langle
S(\delta-h)|S(\delta+h)\rangle \right|=\prod_{i=1}^{N_{\mathrm{s}}}
\left|\vect{S}_i(\delta-h) \cdot \vect{S}_i(\delta+h)\right|$, where we use
$|S(\delta)\rangle$ as a shorthand for the full spin configuration $\{\vect{S}_i(\delta) \}$ at a disorder strength $\delta$, the scalar product
between spins $\vect{S}_i$ is the usual vector scalar product and we rotate the
spin configurations $\{ \vect{S}_i\}$ such that $\vect{S}_1$ coincides between
both. Defined in this way we expect the fidelity to change quadratically in $h$ for
small $h$ and be exponential in the number of spins $N_s$. 

The state fidelity is well established as a diagnostic for phase transitions in
quantum systems \cite{Fidelity_Zhou_2008,Fidelity_Gu_2010}. However, for
classical systems the choice of the scalar product is somewhat arbitrary. We
emphasise that here we mainly use it in the basic analysis sense, where we view
the classical ground states as functions of a parameter $\delta$ and simply ask
for continuity or differentiability of this function.

\begin{figure}
\begin{minipage}{0.99\columnwidth}
  \includegraphics[width=.99\columnwidth]{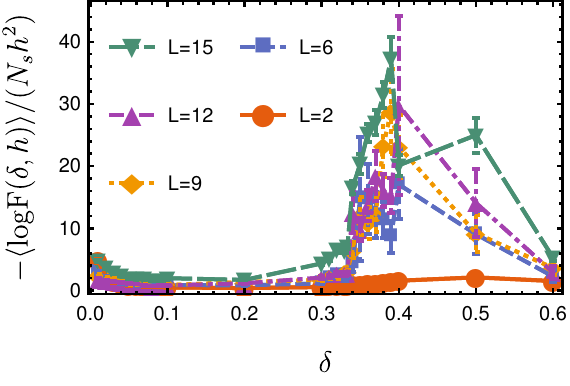}
\end{minipage}
   \caption{Logarithmic fidelity per site $\log F(\delta,h)/N_s$ for $h=0.001
     \delta$ of the groundstate at $T=0$.
 \label{fig:fidelity}}
\end{figure}

In Fig.~\ref{fig:fidelity} we show the disorder average of the logarithmic
fidelity normalised to the number of sites $\langle \log
F(\delta,h)/(N_{\mathrm{s}} h^2)\rangle$ as a function of disorder strength. 
The fidelity clearly tracks the phase-transition at $\delta_c=1/3$ where the
classical state changes rapidly for small changes of the system parameters.
In addition, we observe a small peak in the low delta regime, which however
scales to 0 for larger system sizes, whereas the peak at the transition scales
with $N_{\mathrm{s}}$.

This suggests that the states connect smoothly to the ground states of the
clean kagome system in the limit $\delta \rightarrow 0$ and evolve smoothly up
to the critical point at $\delta_c$.
In the next section we provide semi-analytic arguments that support this picture.

\sectionn{Continuity of states and implicit function theorem}
We may understand the evolution of the classical ground states of the model with disorder strength $\delta$ via the mapping
\begin{equation}
  \label{eq:inv_function}
  \begin{split}
  G : \quad &\mathbb{R} \times \mathbb{R}^{3 N_{\mathrm{s}}}\rightarrow \mathbb{R}^{3N_{\mathrm{s}}} \\
      & \delta \times \{ \vect{S}_i\} \mapsto
        \begin{cases}
          \vect{S}^2_i-1  & i \in 1,\dots,N_S \\
          \bL_{\alpha} & \alpha \in 1, \dots, 2 N_{\mathrm{s}}/3
        \end{cases}
        \end{split}
\end{equation}
where the dependence of $\bL$ on the spins and $\delta$ is implicit.
The ground state configurations then correspond to the preimage of the
zero-vector, e.g. $\{ \vect{S}^{\mathrm{gs}}_i\} = G^{-1}(\vect{0})$.

Given a ground state at some fixed disorder strength, e.g.
a point $\{\delta_0,\{\vect{S}_i\}\}$ such that
$G(\{\delta_0,\{\vect{S}_i\}\})=\vect{0}$, the implicit function theorem
guarantees that the ground state is given by a differentiable function of the
disorder strength $\delta$ in an open neighbourhood of $\delta_o$ if the Jacobian
$\left[ \frac{\partial G_i}{ \partial S_{j d}} \right]$ is invertible. Here $j=1,\dots,N_{\mathrm{s}}$
is the site index and $d=x,y,z$ is the index of the spatial dimension.

Strictly speaking one needs to consider this mapping on the quotient space $\mathbb{R}^{3N_{\mathrm{s}}}/\mathrm{O}(3)$
to remove the (trivial) degeneracy due to global $\mathrm{O}(3)$ rotations. This
can be done in different ways, e.g. by fixing one spin and one plane and
considering the remaining coordinates. We find it more convenient simply to
suppress the three zero singular values of the Jacobian corresponding to
this degeneracy.

The implicit function theorem ensures both existence and the smooth
dependence on disorder strength of the ground states, at least in some
neighbourhood of a non-singular point.
In particular, if all ground states are non-singular the number of ground states
is also preserved when increasing the disorder strength. Further, when during
this mapping one does not encounter a singular point, one can map all states back to ground states of the clean model at $\delta=0$, or starting
from these obtain all ground states at finite disorder.

Based on the form of $G$ in Eq.~\ref{eq:inv_function} and its Jacobian one can
already make some important observations:
Firstly, for coplanar states the Jacobian is necessarily singular, thus, we do
not expect coplanar states to connect to finite disorder ground states.
Secondly, the Jacobian is also singular if two spins in a triangle are
collinear. Consequently, as soon as defect triangles appear at $\delta_c$, the
mapping based on the implicit function theorem breaks down.

\begin{figure}
\begin{minipage}{\columnwidth}
  \includegraphics[width=.99\columnwidth]{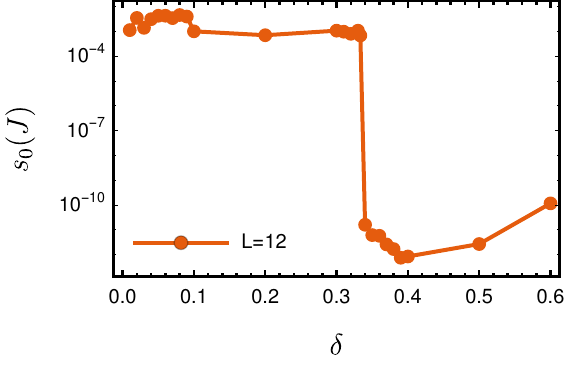}
\end{minipage}
 \caption{Lowest singular value $s_0$ (suppresing the three zeros due to global
   rotations) of the Jacobian of $G$, Eq.~\ref{eq:inv_function}, found over
   different ground states ($T=0$) and disorder realisations at a strength $\delta$. This indicates
   whether the ground state may be continued locally via the inverse function
   theorem as $\delta < \delta_c$ is varied.
 \label{fig:jacobian_singular_values}}
\end{figure}

To test whether the ground states we find actually are non-singular, we consider
the lowest singular value of the Jacobian (suppressing the 3 zeros due to global
rotation) for ground states found at different disorder configurations and
strengths. In Fig. ~\ref{fig:jacobian_singular_values} we show the lowest such
value found over 1000 disorder realisations and 20 states for every point.

We clearly observe the transition at $\delta_c=1/3$ as explained by the
appearance of defect triangles. Further, we find no singular states below the
transition. Thus, we expect all ground states of the disordered system to connect smoothly to non-coplanar
ground states of the clean system.

\sectionn{Fate of coplanarity}
We consider the stability of coplanar states to disorder and establish that they
are not part of the ground state manifold of the disordered models.

\begin{figure}
\begin{minipage}{0.99\columnwidth}
  \includegraphics[width=.99\columnwidth]{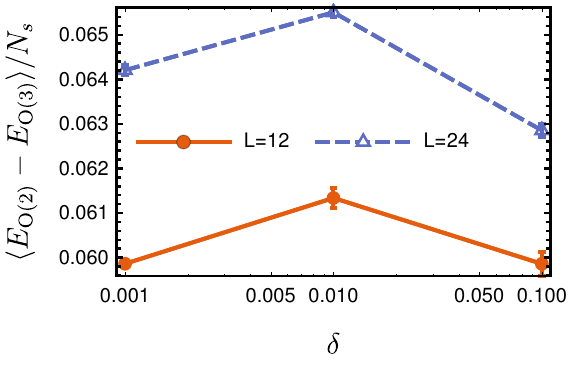}
\end{minipage}
\caption{Energy difference per spin of ground states ($T=0$) of n-component spins
   $E_{\mathrm{O}(n)}$ for $n=3$ (Heisenberg) and $n=2$ (XY) as a function of
   disorder strength $\delta$ for linear system sizes $L=12,24$ with total
   number of spins $N_{\mathrm{s}}=3 L^2$ averaged over disorder realisations.
 \label{fig:XY_comparison}}
\end{figure}

To do so we compare the minimal energy of spin configurations of 2 and 3
component spins respectively obtained by numerical optimisation and averaged
over disorder realisations.
The results for different disorder strengths and linear system sizes are shown
in Fig.~\ref{fig:XY_comparison}.
We observe that the coplanar $\mathrm{O}(2)$ ground states always have a
higher energy than the corresponding $\mathrm{O}(3)$ state. We emphasise that
this is in fact true individually for all disorder realisations and not only for
the mean.
Moreover, this energy difference increases with increasing system size, and
consequently would appear to remain finite in the thermodynamic limit.

We conclude that as the coplanar states have higher energy than the non-coplanar
states entropic selection of coplanar states should not occur for the disordered
model in contrast to the clean kagome antiferromagnet. This is consistent with
our results on the heat capacity in the main text.

\begin{figure}
  \begin{minipage}{0.99\columnwidth}
   \includegraphics[width=.99\columnwidth]{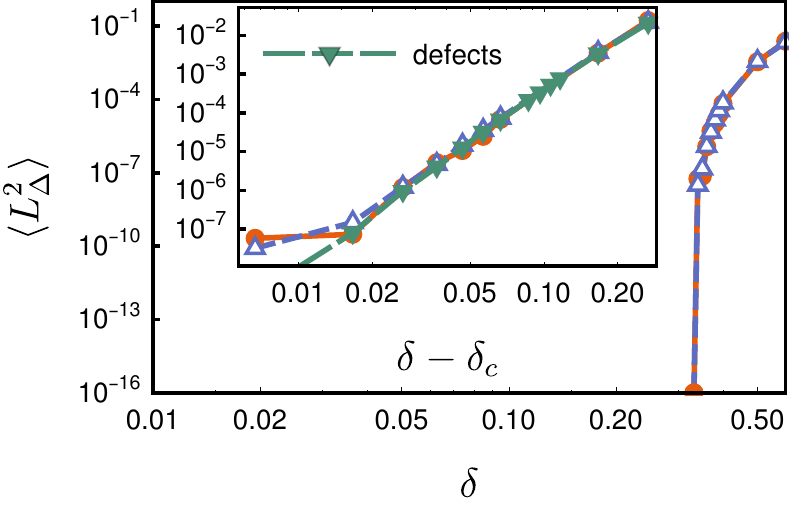}
 \end{minipage}
 \caption{Residual energy per triangle of the MCM vs.\ $\delta$ for $L=12,24$
   (circles, triangles) on a log-log scale.
  % Statistical errors are smaller than symbol size.
   Inset: Close up close to the critical point compared to the direct evaluation of energy of "defect" triangles with collinear spins.
 \label{fig:gs_MCM_L0}}
\end{figure}

\sectionn{Residual energy of the MCM}
In this section we provide the residual energy of the MCM as a function of
disorder strength $\delta$ (the corresponding plot for the BDM can be found in
the main text).
In Fig.~\ref{fig:gs_MCM_L0} we clearly observe the sharp transition at
$\delta_c=1/3$ below which the residual energy per triangle $\langle
\vect{L}^2_{\Delta} \rangle$ is strictly zero (within machine precision).
In the inset we show a zoom into the behaviour close to the critical point.
We observe the same scaling $\langle \vect{L}^2_{\Delta} \rangle \sim
(\delta-\delta_c)^5$ as for the BDM.

In addition, we again compare the residual
energies obtained from the direct numerical optimisation with the prediction of
``defects''. This estimate is obtained by individually minimising the energy
independently on all triangles. As explained above for a single triangle, the
minimal energy is zero if $\gamma_1 \ge \gamma_2+\gamma_3$, and
$\gamma_1-(\gamma_2+\gamma_3)$ otherwise, where we assume an ordering
$\gamma_1 \ge \gamma_2\ge\gamma_3$. The later case we refer to as ``defect triangles''
and these are the only ones that contribute to this estimate.
We observe that this estimate seems to capture the energy of the full connected
system quite well.

\sectionn{Comparison of the correlations}
\begin{figure}
\begin{minipage}{\columnwidth}
  \includegraphics[width=.99\columnwidth]{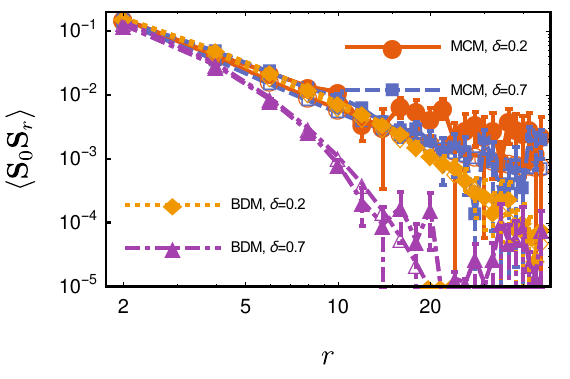}
\end{minipage}
 \caption{Comparison of the spin correlations in the ground state ensemble of $O(3)$
   spins (filled symbols) to the self-consistent Gaussian apprximation (open
   symbols) at $T=0$ for both the bond-disordered model (BDM) and the
   maximal Coulomb model (MCM) in a system of linear size $L=48$ and
   disorder strengths $\delta=0.2,0.7$.
 \label{fig:cors_comparison}}
\end{figure}
We next provide a comparison of the correlations in the ground state ensemble of
$O(3)$-spins to the results of the self-consistent gaussian approximation (SCGA)
for both the BDM and the MCM.

Fig.~\ref{fig:cors_comparison} shows the results for values
of the disorder strength below the transition for $\delta=0.2$ and
above for $\delta=0.7$ in a system with linear size $L=48$.
We observe good agreement for both models and values of $\delta$ within
the statistical errors of the ground state calculation.
In addition, the correlations at large distances are seen to be exponentially suppressed for
the BDM and decay algebraically for the MCM.
Finally, we emphasise that the results of the SCGA have considerably less
statistical noise and allow the study of larger system sizes as exploited in the
main text.

\sectionn{Magnetic structure factor}
Here, we compare the static magnetic structure factor of the BDM to the disorder-free
system obtained from MC simulations at finite temperatures.

\begin{figure}
  \begin{minipage}{0.49 \columnwidth}
\raggedleft (a)
\raggedright \includegraphics[width=.99\columnwidth]{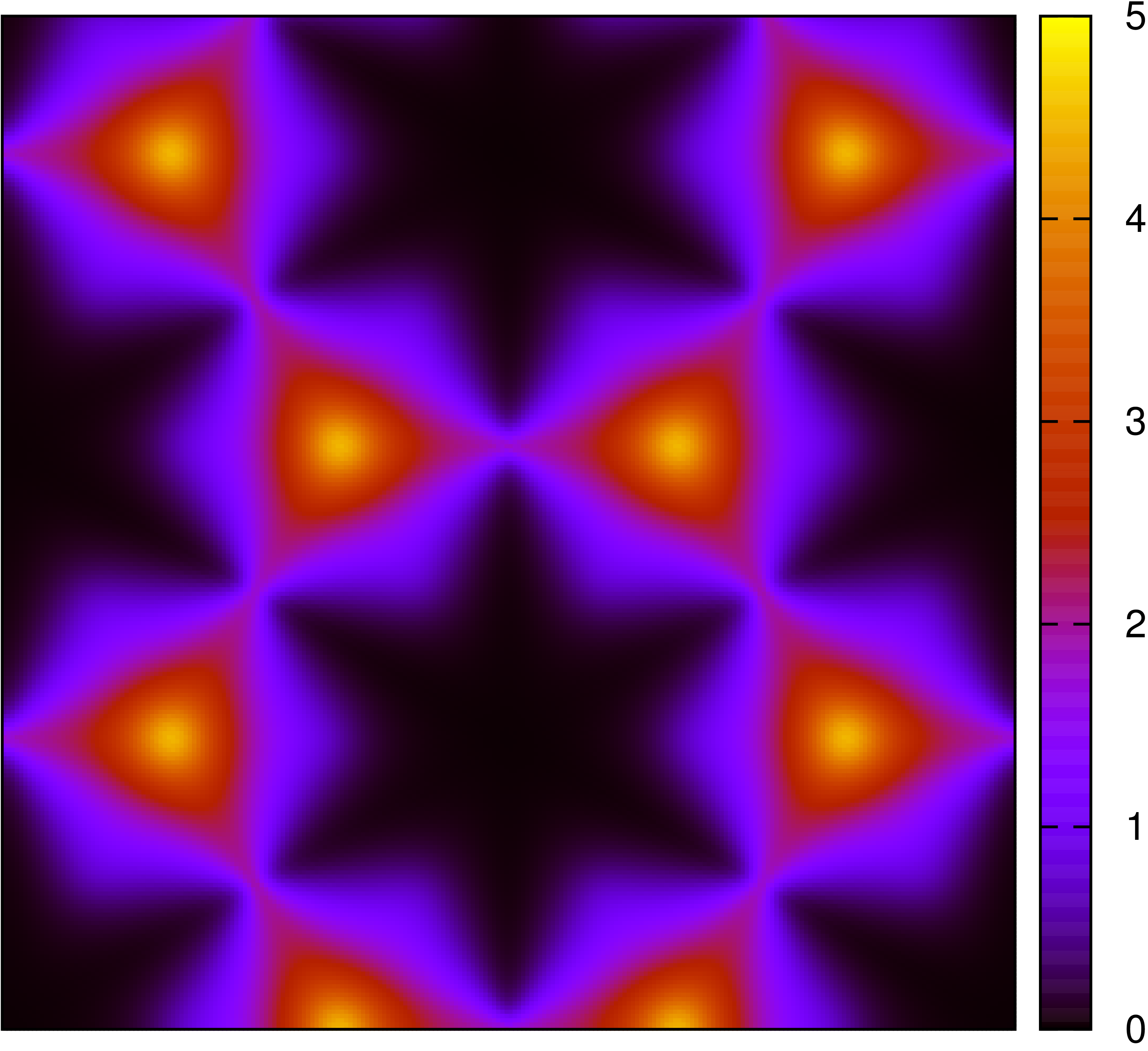}
  \end{minipage}
 \begin{minipage}{0.49 \columnwidth}
 \raggedleft (b)
 \raggedright \includegraphics[width=.99\columnwidth]{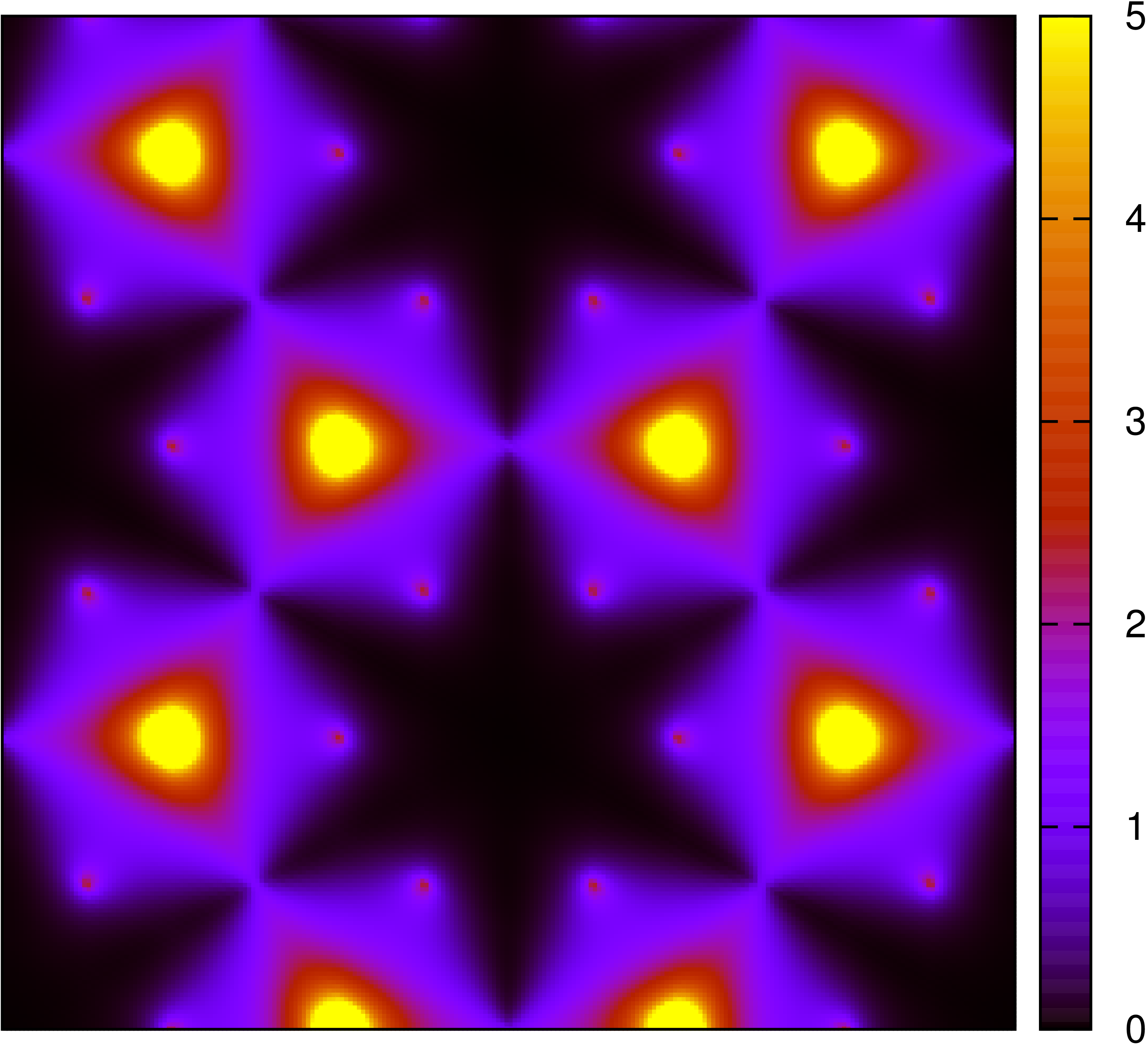}
  \end{minipage}
  \begin{minipage}{0.49 \columnwidth}
  \raggedleft(c)
  \raggedright\includegraphics[width=.99\columnwidth]{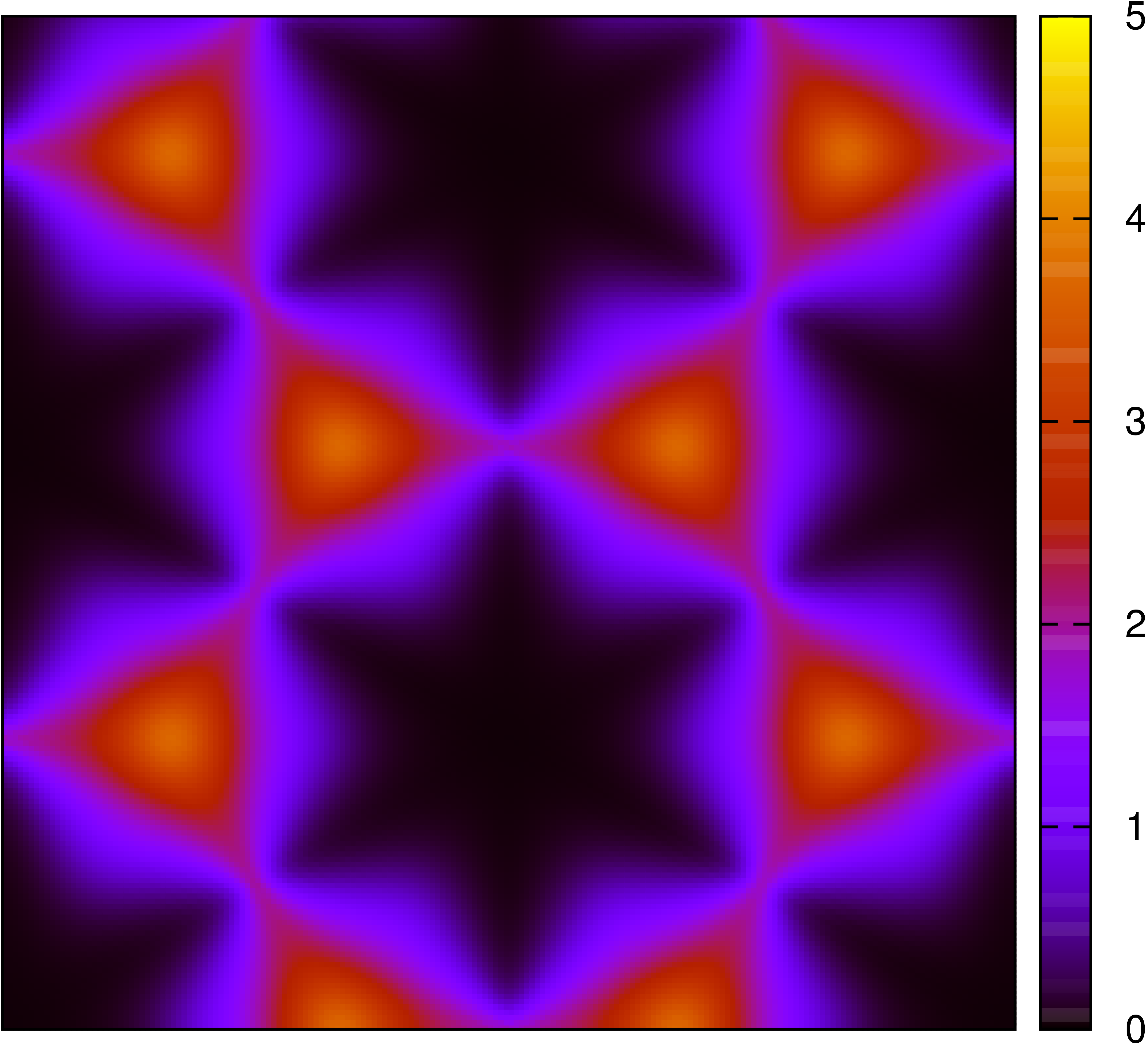}
  \end{minipage}
  \begin{minipage}{0.49 \columnwidth}
  \raggedleft (d)
  \raggedright\includegraphics[width=.99\columnwidth]{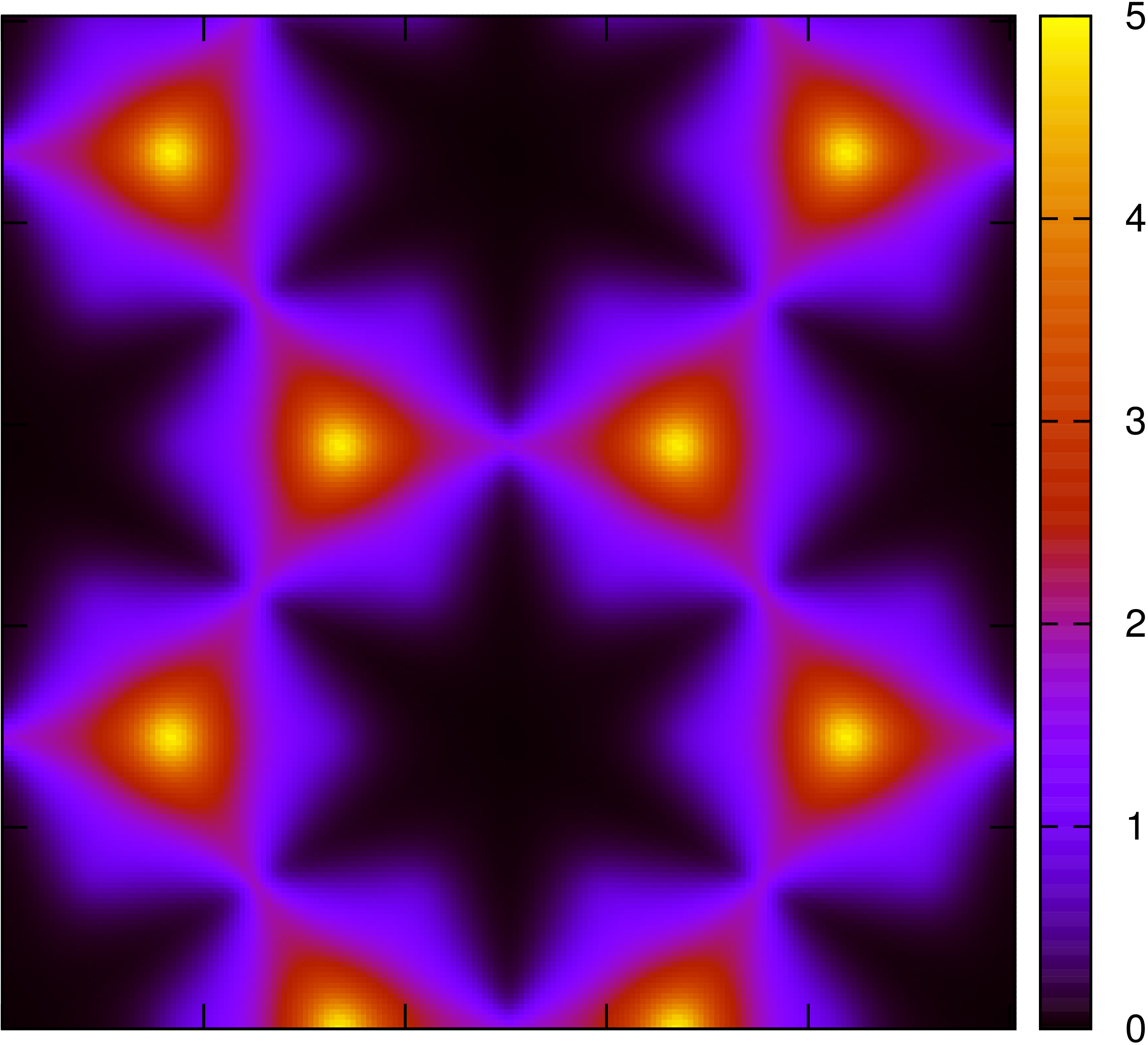}
  \end{minipage}
  \caption{Comparison of the magnetic structure factor of the BDM to
    disorder-free kagome antiferromagnet in different temperature regimes.
    Region in the momentum space corresponds to $0\leq q_{x,y} \leq 8 \pi$.
    Temperature in the top and bottom row $T/J=0.005$ and $T/J=0.02$
    respectively.
    Left column for the BDM at disorder strength $\delta=0.2$,
    right column for the disorder-free model. \label{fig:MC_structure_factor}}
 \end{figure}

In Fig.~\ref{fig:MC_structure_factor} we observe that the BDM (panel a) does not develop the additional $\sqrt{3}\times \sqrt{3}$
peaks at low temperatures present for the disorder-free case (panel b).
In addition, the BDM at a disorder strength $\delta$ with exponentially screened
correlations compares well to the disorder-free case at a finite temperature
$T^{*} \sim \delta^2$ which also exhibits screened correlations due to thermal
fluctuations. This is demonstrated by the top left (panel a) for the BDM at
$\delta=0.2$ and the bottom right (panel d) for the disorder free case at $T/J=0.02$.

 \sectionn{Spin glass transition}
\begin{figure}
  \begin{minipage}{\columnwidth}
    \includegraphics[width=.99\columnwidth]{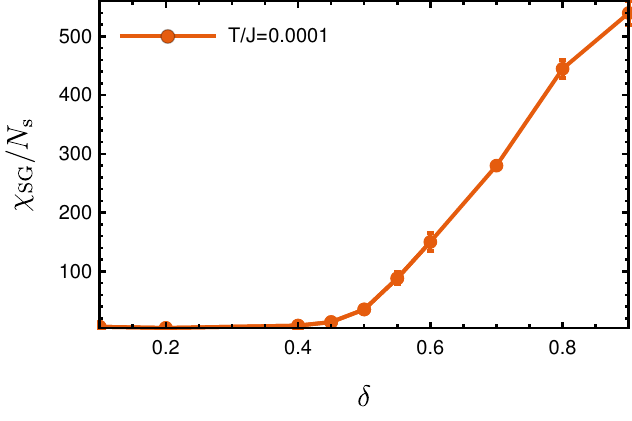}
  \end{minipage}
   \caption{Extrapolated spin glass susceptibility $\chi_{\mathrm{SG}}$ vs. disorder strength
     for $T/J=10^{-4}$ from Monte-Carlo simulations for the BDM.
     \label{fig:MC_results2}}
 \end{figure}
Fig.~\ref{fig:MC_results2} shows the Monte-Carlo results for the spin glass susceptibility extrapolated to the
thermodynamic limit based on systems with $L=4,\cdots,30$ at a fixed temperature
$T/J=10^{-4}$ versus the disorder strength $\delta$.
It clearly demonstrates the absence of spin glass correlations in the
jammed-spin liquid and the existence of a spin glass for large $\delta$.
However, we did not precisely determine the exact transition point into the spin
glass $\delta_{\mathrm{SG}}$.
In particular, we do not exclude the presence of an intermediate phase for
$\delta_c<\delta<\delta_{SG}$.

\section{Checkerboard/planar pyrochlore lattice}
In this section we provide a discussion of the physics of our model defined on
the checkerboard/planar pyrochlore lattice for O$(2)$ spins.

We emphasise that in contrast to the kagome lattice on the planar pyrochlore
lattice the bond-disordered model Eq.(1) (main text) is not equivalent to
the model defined via scaling factors Eq.(2) (main text) as the former has 3 independent
degrees of freedoms per spin (3 bond couplings per spin) and the latter only two
scaling factors per spin. Thus, for a generic bond-disordered model it is not possible to write
it as a sum of squares as for the kagome lattice.
%The case of true bond-disorder has been discussed in the literature (cite CHALKER)
%and has a spin-glass transition for any disorder strength.

\begin{figure}
  \begin{minipage}[t]{0.49 \columnwidth}
\raggedleft (a)
\raggedright \includegraphics[width=.99\columnwidth]{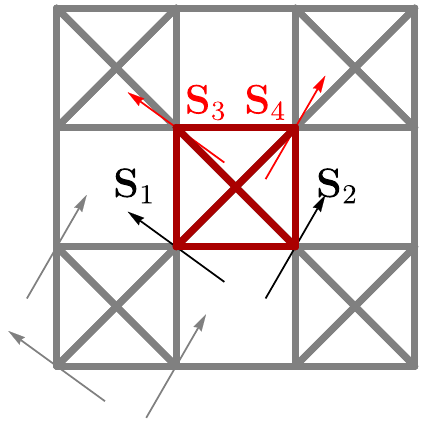}
  \end{minipage}
 \begin{minipage}[t]{0.49 \columnwidth}
 \raggedleft (b)
\raggedright \includegraphics[width=.99\columnwidth]{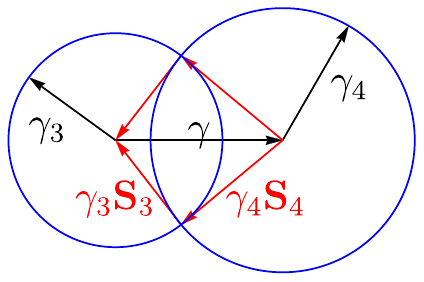}
  \end{minipage}
  \caption{(a) Planar pyrochlore lattice, gray and black spins
    ($\vect{S}_1$, $\vect{S}_2$) are assumed
    known, red spins $\vect{S}_3$, $\vect{S}_4$ are to be determined.
    (b) Groundstate constraint in the bold central square, $\gamma = \left|\gamma_1
      \vect{S}_1+\gamma_2 \vect{S}_2  \right| $, allows generically two
solutions for $\vect{S}_3$ and $\vect{S}_4$.
\label{fig:planar_pyrochlore}}
\end{figure}

Therefore, we will mainly discuss the MCM model in the
following,

\begin{equation}
  H=\frac{1}{2} \sum_\alpha \bL_\alpha^2  ,\quad \text{with} \quad \bL_\alpha=\sum_{i\in\alpha}\gamma_{i\alpha}\bs_i \, ,
\label{eq:app_H_pyro}
\end{equation}
where $\alpha$ now denotes the fully connected squares of the planar pyrochlore lattice
illustrated in Fig.~\ref{fig:planar_pyrochlore}.

\sectionn{Analysis of a single square}
We provide an estimate of the expected critical point based on the analysis of
the constraint on a single square of the planar pyrochlore lattice.

The constraint
\begin{equation}
  0=\sum_{i\in \Box} \gamma_i \vect{S}_i= \sum \gamma_1
  \vect{S}_1+\gamma_2 \vect{S}_2+\gamma_3 \vect{S}_3 +\gamma_4 \vect{S}_4
\end{equation}
clearly becomes unsatisfiable when $\gamma_1 > \gamma_2+\gamma_3+\gamma_4$,
where we assume the $\gamma_i$ to be ordered in magnitude. If we again choose
$\gamma \in (1-\delta, 1+\delta)$, and consider the extremal case
$\gamma_1=1+\delta$, $\gamma_2=\gamma_3=\gamma_4=1-\delta$, this implies $\delta_c = 1/2$.

\sectionn{Transfer matrix}
We continue with the transfer matrix argument as
applicable to the planar pyrochlore lattice.

In the arrangement shown in Fig.~\ref{fig:planar_pyrochlore}(a) we have already
chosen all spins to the left and below the black square of the planar
pyrochlore lattice containing the known spins $\vect{S}_1$ and $\vect{S}_2$
and the unknown spins $\vect{S}_3$ and $\vect{S}_4$.
We may rewrite the groundstate constraint in this
square as
\begin{equation}
  0=\sum_{i\in \Box} \gamma_i \vect{S}_i= \sum (\gamma_1
  \vect{S}_1+\gamma_2 \vect{S}_2)+\gamma_3 \vect{S}_3 +\gamma_4 \vect{S}_4
\end{equation}
to recognise that it is equivalent to demanding that the vectors $(\gamma_1
\vect{S}_1+\gamma_2 \vect{S}_2)$, $\gamma_3 \vect{S}_3$, $\gamma_4 \vect{S}_4$
form a closed triangle. As a triangle with three side lengths known is uniquely
determined up to orientation, and we already know the orientation of one side,
there remains a discrete choice between two configurations, related by
mirror-reflection along the known side. We may then repeat this step to
determine all spins in the next layer, and then throughout all layers to determine all spins in the lattice.

Ignoring potentially inconsistent configurations this yields an estimate of the
number of groundstates $N_{gs} \sim 2 ^{N_s/2}$ as in every step we have 2
choices to fix 2 spins.

\sectionn{Constraint satisfaction}
In this section we provide numerical evidence that the groundstates on the
connected planar pyrochlore lattice indeed exactly satisfy the groundstate
constraint up to the critical disorder strength $\delta_c=1/2$. In addition, we
contrast these results to the truly bond-disordered model to show that the requirement
that the model may be written as a sum of squares is indeed crucial to the existence of
a ``jammed'' phase.

\begin{figure}
  \begin{minipage}{0.99 \columnwidth}
 \includegraphics[width=.99\columnwidth]{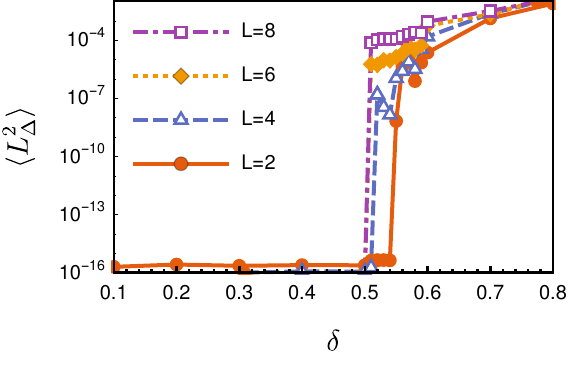}
  \end{minipage}
  \caption{Residual energy $\langle  L^2_{\Box} \rangle$ per square versus
    disorder strength for the MCM, Eq.~\ref{eq:app_H_pyro}, on the planar pyrochlore lattice for square systems with $L_x=L_y=2,4,6,8$
\label{fig:pyro_L0}}
\end{figure}

\begin{figure}
  \begin{minipage}{0.99 \columnwidth}
 \includegraphics[width=.99\columnwidth]{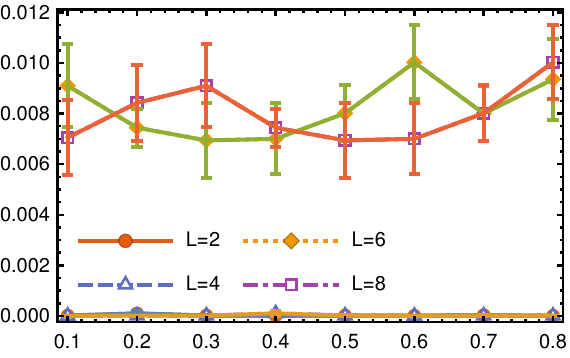}
  \end{minipage}
  \caption{Residual energy $\langle  L^2_{\Box} \rangle$ per square versus
    disorder strength for the bond-disordered model on the planar pyrochlore lattice for square systems with $L_x=L_y=2,4,6,8$
\label{fig:pyro_BD_L0}}
\end{figure}
We show the resulting residual energies in Fig.~\ref{fig:pyro_L0}. The results
clearly demonstrate the transition at $\delta_c=1/2$ as expected from the analysis of
an isolated square above. We emphasise that we expect the transition to also be continuous, and the jump only occurs as the lowest disorder strength simulated above the
critical point is $\delta=0.51$, for which we expect a residual energy on the
order of $\langle  L^2_{\Box} \rangle \sim 10^{-5}$.

We may contrast this behaviour to that of the bond-disordered model shown in
Fig.~\ref{fig:pyro_BD_L0}. Here we see a finite non-vanishing residual energy essentially
independent of $\delta$ in the considered regime and there is no indication of
any transition or a ''jammed `` phase in this model.

\bibliographystyle{plain}
%\bibliography{kagome_bib}{}
\input{supplemental.bbl}

\end{document}